\documentstyle[aps,preprint]{revtex}
\input epsf
\global\firstfigfalse  

\tighten

\begin{document}
\pagestyle{empty}

\preprint{
\begin{minipage}[t]{3in}
\end{minipage}
}

\title{Coupled Quintessence in a Power-Law Case and the Cosmic Coincidence
Problem }

\author{Xin Zhang \\\bigskip}
\address{Institute of High Energy Physics, Chinese Academy of Sciences\\
P.O.Box 918(4), Beijing 100049, People's Republic of China\\
\smallskip{\tt zhangxin@mail.ihep.ac.cn}}

\maketitle

\begin{abstract}

The problem of the cosmic coincidence is a longstanding puzzle.
This conundrum may be solved by introducing a coupling between the
two dark sectors. In this Letter, we study a coupled quintessence
scenario in which the scalar field evolves in a power law
potential and the mass of dark matter particles depends on a power
law function of $\phi$. It is shown that this scenario has a
stable attractor solution and can thus provide a natural solution
to the cosmic coincidence problem.

\vskip .4cm


\vskip .2cm


\end{abstract}

\newpage
\pagestyle{plain} \narrowtext \baselineskip=18pt

\setcounter{footnote}{0}

Before the accelerated expansion of the universe was revealed by
high red-shift supernovae Ia (SNe Ia) observations \cite{sn}, it
could hardly be presumed that the main ingredients of the universe
are dark sectors. The concept of dark energy was proposed for
understanding this currently accelerating expansion of the
universe, and then its existence was confirmed by several high
precision observational experiments \cite{wmap,sdss}, esp. the
Wilkinson Microwave Anisotropy Probe (WMAP) satellite experiment
\cite{wmap}. The WMAP shows that dark energy (DE) occupies about
$73\%$ of the energy of our universe, and dark matter (DM) about
$23\%$. The usual baryon matter which can be described by our
known particle theory occupies only about $4\%$ of the total
energy of the universe. Although we know that the ultimate fate of
the universe is determined by the nature of DE, the information
about its nature we can acquire is still very limited. So far the
confirmed information about DE can be summarized as the following
three items: it is a kind of exotic component with negative
pressure such that it can accelerate the universe's  expansion; it
is spatially homogeneous and non-clustering; and it dominates the
universe today although it contributes little to the universe at
the early times.

The investigation of the nature of DE is an important mission in
the modern cosmology. Much work has been done on this issue, and
there is still a long way to go. One candidate for DE is vacuum
energy density or cosmological constant $\Lambda$ \cite{cc} for
which the equation of state $w=-1$. The cosmological model that
consists of a mixture of vacuum energy and cold dark matter (CDM)
is called LCDM (or $\Lambda$CDM). Another possibility is the
so-called QCDM cosmology which based upon a mixture of CDM and
quintessence field \cite{quin}. The energy density and the
negative pressure are provided by the quintessence scalar field
$\phi$ slowly evolving down its potential $V(\phi)$. The equation
of state of the quintessence $-1<w<-1/3$ is guaranteed by the slow
evolution. However, as is well known, there are two difficulties
arise from all of these scenarios, namely, the ``fine-tuning''
problem, and the ``cosmic coincidence'' problem
\cite{coincidence}. The cosmic coincidence problem states: Since
the energy densities of DE and DM scale so differently during the
expansion of the universe, why are they nearly equal today? To get
this coincidence, it appears that their ratio must be set to a
specific, infinitesimal value in the very early universe.

A possible solution to this cosmic coincidence problem may be
provided by introducing a coupling between quintessence DE and
CDM. This coupling is often described by the varible-mass particle
(VAMP) scenario \cite{vamp}. The VAMP scenario assumes that the
CDM particles interact with the scalar DE field resulting in a
time-dependent mass, i.e. the mass of the CDM particles evolves
according to some function of the scalar field $\phi$. It has been
shown that if we choose an exponential potential to the
quintessence scalar field $\phi$ and at the same time assume the
CDM particle mass also depends exponentially on $\phi$, the
late-time ratio between DM energy density $\rho_\chi$ and DE
energy density $\rho_\phi$ will remain constant \cite{couple}.
This behavior relies on the existence of an attractor solution,
which makes the effective equation of state of DE mimic the
effective equation of state of DM at the late times so that the
late time cosmology insensitive to the initial conditions for DE
and DM. In addition, the final effective equation of state of both
components is negative and may lead, if this value is less than
$-1/3$, to an accelerated expansion of the universe. Therefore,
the scenario containing quintessence with exponential potential
and VAMPs with exponential function of $\phi$ solves the cosmic
coincidence problem in this sense. However, it should be pointed
out that this case is not the unique case. We consider here
another case --- the power law case.

In this Letter, we explore the case in which the scalar field
evolves in a power law potential and the mass of VAMPs depends on
a power law function of $\phi$. We find that the power law case
also has a stable attractor solution which is similar to the
exponential case. We numerically solve the equation of motion of
the scalar field $\phi$ and then illustrate the behavior of DE and
DM. The numerical results show that the power law case of coupled
DE with VAMPs also solves the cosmic coincidence problem.

Consider, now, the CDM particle $\chi$ with mass $M$ depending on
a power law function of the DE field $\phi$,
\begin{equation}
M_\chi(\phi)=M_{\ast}\phi^{-\alpha}~,\label{m}
\end{equation}
where $\phi$ is expressed in units of the reduced Planck mass
$M_p$ ($M_p\equiv 1/\sqrt{8\pi G}=2.436\times 10^{18} {\rm GeV}$),
and $\alpha$ is a positive constant. The scalar field has a power
law potential
\begin{equation}
V(\phi)=V_{\ast}\phi^\beta~,\label{v}
\end{equation} where $\beta$ is a positive constant. Since the CDM
particle is stable, its number density $n_{\chi}$ must obey the
equation
\begin{equation}
\dot{n}_\chi+3H n_\chi=0~,\label{n}
\end{equation} where the dot denotes a derivative with respect to
time, $H=\dot{a}/a$ represents the Hubble parameter, and $a(t)$ is
the scale factor of the universe. The energy density of DM
$\rho_\chi$ is also $\phi$-dependent, which is given by
\begin{equation}
\rho_\chi(\phi)=M_\chi(\phi)n_\chi~,\label{rhochi}
\end{equation} and it follows that
\begin{equation}
\dot{\rho}_\chi+3H\rho_\chi=-\alpha{\dot{\phi}\over\phi}\rho_\chi~.\label{eomdm}
\end{equation}
Since the total energy of DE and DM is conserved, the equation of
motion for DE can be obtained
\begin{equation}
\dot{\rho}_{\phi}+3H\rho_\phi(1+w_\phi)=\alpha{\dot{\phi}\over\phi}\rho_\chi~,\label{eomde}
\end{equation} where the usual parameter of equation of state for
the homogeneous scalar field is given by
\begin{equation}
w_\phi={p_\phi\over\rho_\phi}={{1\over
2}\dot{\phi}^2-V(\phi)\over{1\over
2}\dot{\phi}^2+V(\phi)}~.\label{wphi}
\end{equation}

The equations of motion (\ref{eomdm}) and (\ref{eomde}) can also
be written in the form of effective equations of state for DM and
DE
\begin{equation}
\dot{\rho}_{\chi}+3H\rho_\chi(1+w_\chi^{(e)})=0~,\label{effdm}
\end{equation}
\begin{equation}
\dot{\rho}_{\phi}+3H\rho_\phi(1+w_\phi^{(e)})=0~,\label{effde}
\end{equation}
where
\begin{equation}
w_\chi^{(e)}={\alpha\over 3H}{\dot{\phi}\over\phi}={\alpha\over
3}{\phi'\over\phi}~,\label{effwdm}
\end{equation}
\begin{equation}
w_\phi^{(e)}=w_\phi-{\alpha\over
3H}{\dot{\phi}\over\phi}{\rho_\chi\over\rho_\phi}=w_\phi-{\alpha\over
3}{\phi'\over\phi}{\rho_\chi\over\rho_\phi}~,\label{effwdm}
\end{equation} are the effective equation of state parameters for
DM and DE, respectively. Primes denote derivatives with respect to
$u=\ln(a/a_0)=-\ln(1+z)$, where $z$ is the red-shift, and $a_0$
represents the current scale factor. From (\ref{eomde}) or
(\ref{effde}) one can get the equation of motion for the scalar
field $\phi$,
\begin{equation}
\ddot{\phi}+3H\dot{\phi}={\alpha\over\phi}\rho_\chi-{\beta\over\phi}V~.\label{eomphi}
\end{equation}

The Friedmann equation for a spatially flat universe with DE, DM,
baryons, and radiation reads
\begin{equation}
3H^2={1\over 2}\dot{\phi}^2+V(\phi)+\rho_\chi(\phi)+\rho_{\rm
b}+\rho_{\rm rad}~,\label{friedmann}\end{equation} the Planck
normalization $M_p=1$ has been used here. Using the Friedmann
equation (\ref{friedmann}), the equation of motion of field $\phi$
can be written in terms of the derivatives of $\phi$ with respect
to $u$,
\begin{equation}
H^2\phi''+{1\over 2}(\rho_\chi+\rho_{\rm b}+{2\over 3}\rho_{\rm
rad}+2V)\phi'={\alpha\over\phi}\rho_\chi-{\beta\over\phi}V~,\label{eom}
\end{equation}
where
\begin{equation}
H^2={1\over 3}{(\rho_\chi+\rho_{\rm b}+\rho_{\rm rad}+V)\over
1-\phi'^2/6}~.\label{h2}
\end{equation}

\vskip.8cm
\begin{figure}
\begin{center}
\leavevmode \epsfbox{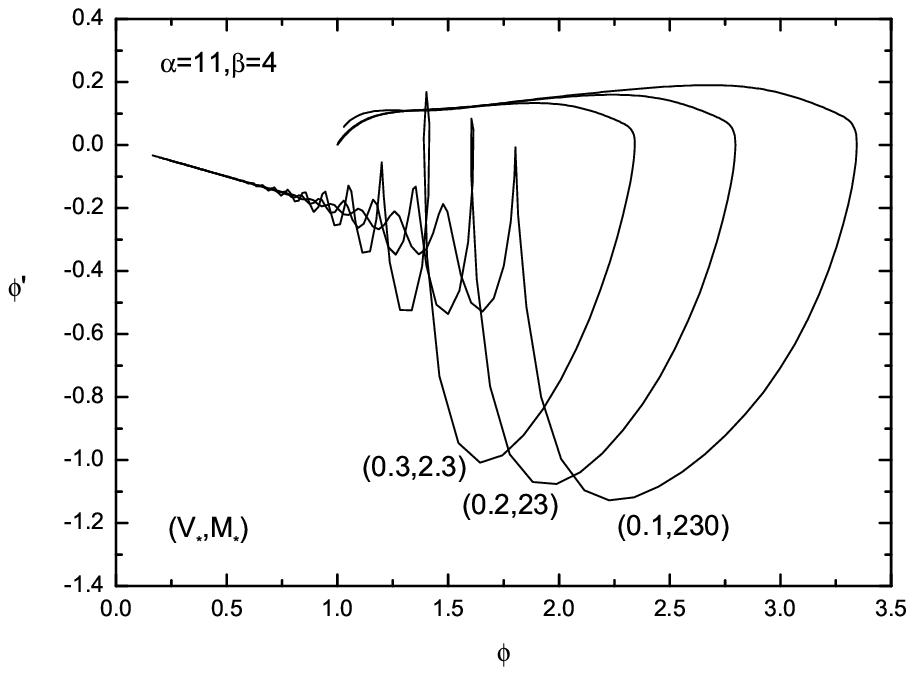} \caption[]{The phase plane for
$\alpha=11$ and $\beta=4$. The units of $V_\ast$ and $M_\ast$ are
in $\rho_{c0}$ and $\rho_{c0}/n_{\chi 0}$, respectively. The three
lines correspond to cases $(V_\ast,M_\ast)$ taken to be (0.1,230),
(0.2,23) and (0.3,2.3), respectively.}
\end{center}
\end{figure}

It is known that the attractor solution can be given analytically
in the exponential case. In this case, however, it is difficult to
obtain the analytical solution of the attractor. The reason arises
from that the late-time behavior of the field $\phi$, unlike in
the exponential case, is not linear in this power law case. In
what follows we will show this difficulty. At the late times, the
energy densities of baryons and radiation are red-shifted away
completely, there remains only DE component and DM component if
the attractor exists. This requires $w_\phi^{(e)}=w_\chi^{(e)}$ in
this limit, it follows that
\begin{equation}
\Omega_\phi\bigg\vert_{a\rightarrow\infty}={1\over3}\phi'^2-{\alpha\over3}{\phi'\over\phi}~.\label{attr}
\end{equation}
The attractor makes the ratio between energy densities of DE and
DM be a constant, i.e. $\Omega_\phi '=0$, which gives
\begin{equation}
2\phi^2\phi'\phi''-\alpha\phi\phi''+\alpha\phi'^2=0~.
\end{equation} We see explicitly that the situation in the case of power law
is more complicated than in the case of exponential. In the
exponential case, $\phi'={\rm const.}$, then using the Friedmann
equation this constant can be determined, furthermore it can be
shown that $\Omega_\phi$ and $\Omega_\chi$ are both
$\phi$-independent constant.\footnote{As an example, we take
$M_\chi(\phi)=M_\ast e^{-\lambda\phi}$, and $V(\phi)=V_\ast
e^{q\phi}$, then the attractor solution is given by
$\phi=\phi_0+{-3\over \lambda+q}u~$,
$\Omega_\phi=1-\Omega_\chi={3+\lambda(\lambda+q)\over(\lambda+q)^2}$.
It can be seen that the attractor depends only on $\lambda$ and
$q$. } However, in this case, since $\phi'\neq {\rm const.}$, we
can not identify the attractor easily. We are now required to
solve the equation of motion of $\phi$ (\ref{eom}) numerically.
Whether the attractor solution exists or not can be confirmed by
the numerical results.

\vskip.8cm
\begin{figure}
\begin{center}
\leavevmode \epsfbox{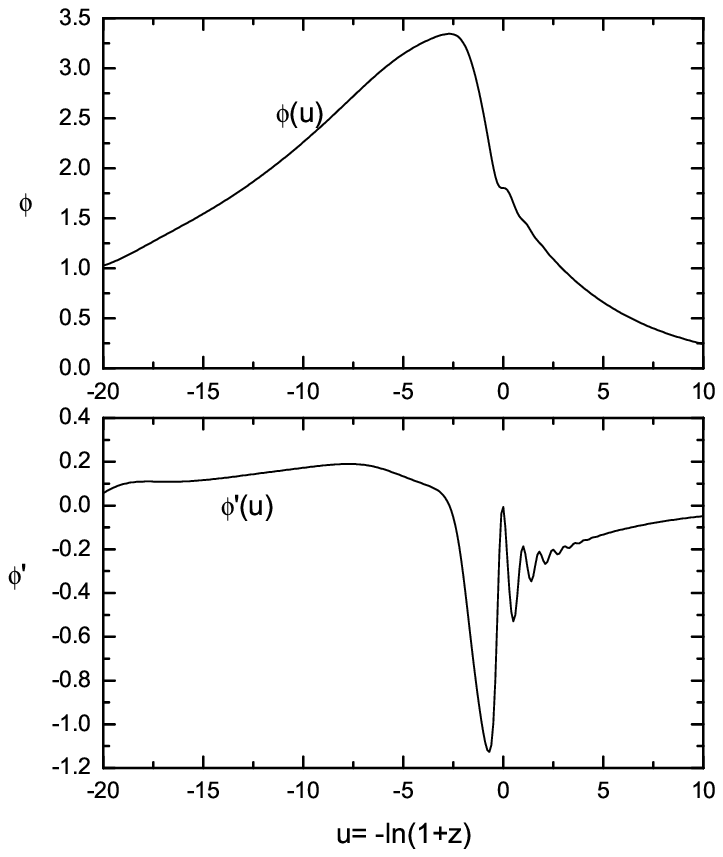} \caption[]{Top panel: A typical
solution for the differential equation (\ref{eom}), namely the
evolution of the $\phi$ field. The corresponding parameter
configuration is: $\alpha=11$, $\beta=4$, $V_\ast=0.1\rho_{c0}$,
and $M_\ast=230\rho_{c0}/n_{\chi0}$. Bottom Panel: the evolution
of $\phi'$ for the same parameters used in top panel.}
\end{center}
\end{figure}

The numerical results indeed show that there exists a stable
attractor solution which depends only on the parameters $\alpha$
and $\beta$ while is very insensitive to the initial conditions
and the chosen values of $V_\ast$ and $M_\ast$. As an example, we
take the case $\alpha=11$ and $\beta=4$. The $(\phi,\phi')$ phase
diagram is plotted in Fig.1. It is shown in the phase plane that
lines corresponding to different chosen values of
$(V_\ast,M_\ast)$ will converge together to the attractor solution
with the cosmological evolution. Note that we express $V_\ast$ and
$M_\ast$ in units of $\rho_{c0}$ and $\rho_{c0}/n_{\chi0}$,
respectively, where $\rho_{c0}=4.2\times 10^{-47}{\rm GeV^4}$ is
the present critical density of universe. It is of interest to
notice that the lines in the phase plane will undergo a period of
oscillation before they converge to the attractor. From Fig.1, we
see that in the attractor region $\phi$ and $\phi'$ both tend to 0
and at the same time $\phi'/\phi$ becomes a constant. Using the
fact that in the attractor region $dV_{eff}/d\phi\sim 0$, where
$V_{eff}=V+\rho_\chi$ is the effective potential, one gets $\alpha
\rho_\chi\simeq\beta V$ in this limit. We re-express this
condition as
\begin{equation}
\alpha\Omega_\chi\simeq\beta(\Omega_\phi-{1\over 6}\phi'^2)~.
\end{equation} Considering $\Omega_\chi=1-\Omega_\phi$ and
$\phi'\sim 0$, one obtains
\begin{equation}
\Omega_\phi\simeq{\alpha\over \alpha+\beta}~.
\end{equation}
Combining with (\ref{attr}), we get
\begin{equation}
{\phi'\over\phi}\simeq -{3\over \alpha+\beta}~,
\end{equation} consequently the attractor solution in the field
space can be expressed clearly as
\begin{equation}
\phi=\phi_0e^{-{3\over\alpha+\beta}u}~,
\end{equation} such that
\begin{equation}
w_\phi^{(e)}=w_\chi^{(e)}= -{\alpha\over\alpha+\beta}~.
\end{equation} The existence of the attractor solution indicates that the role
the coupled quintessence model in a power law case plays in
solving the cosmic coincidence problem is similar to that of the
model in an exponential case.

\vskip.8cm
\begin{figure}
\begin{center}
\leavevmode \epsfbox{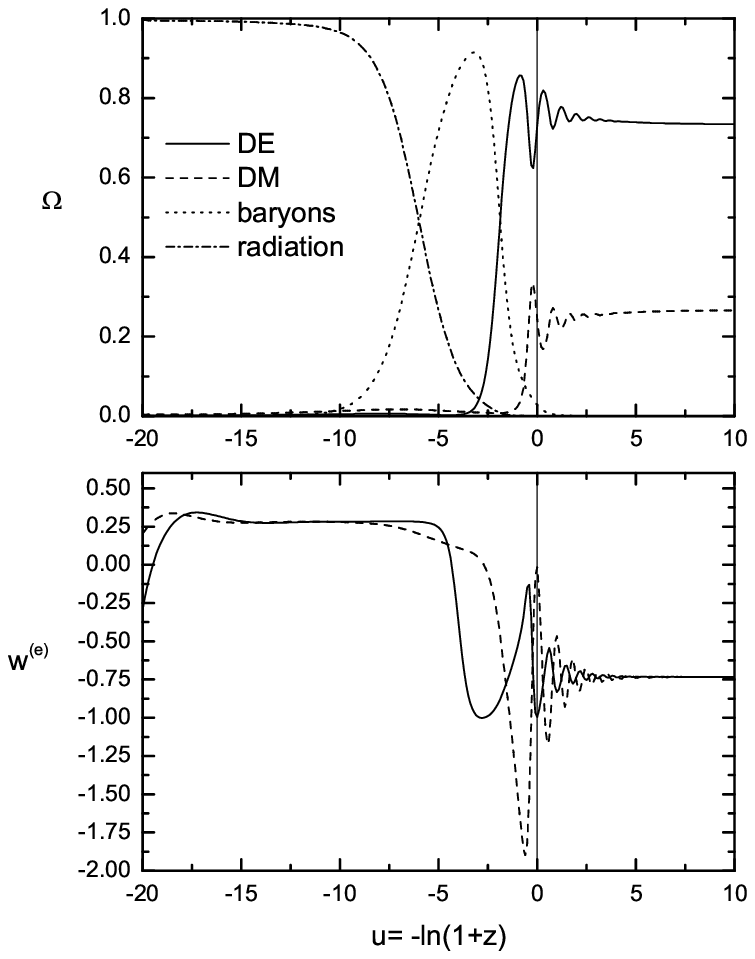} \caption[]{Top panel: The
evolution of the relative abundance of different species,
expressed as fractions of the critical density. The corresponding
model parameters are: $\alpha=11$, $\beta=4$,
$V_\ast=0.1\rho_{c0}$, and $M_\ast=230\rho_{c0}/n_{\chi0}$. Bottom
panel: Effective equations of state for DE (solid line) and DM
(dashed line) for the same parameters used in top panel.}
\end{center}
\end{figure}

We now illustrate the cosmological consequence of this scenario.
The solution of the differential equation (\ref{eom}), namely the
evolution of $\phi(u)$, is plotted in the top panel of Fig.2. Note
that for concreteness the values of $(V_\ast,M_\ast)$ are taken as
$V_\ast=0.1\rho_{c0}$ and $M_\ast=230\rho_{c0}/n_{\chi 0}$. We see
explicitly in this case that $\phi$ increases in the beginning and
begins to decrease around $u\sim -2.5$, which implies that the
$\phi$ field climbs slowly up the potential initially and then
rolls down with the evolution of the universe. Moreover, we notice
that around $u\sim 0$ some wiggle appears. To make the picture of
the kinematics of the field more clear, we plot the 'velocity' of
the field $\phi'(u)$ (note that $\phi'=\dot{\phi}/H$) versus
$u=-\ln(1+z)$ in the bottom panel of Fig.2. It is shown that the
wiggle in the $\phi$-evolution diagram corresponds to the rapid
oscillation of the rolling-down velocity. After the period of
oscillation the field system will enter a stable attractor regime.

It is remarkable that the feature of the coupled quintessence
model is quite different from the scalar DE model in the absence
of the interaction with DM. In Fig.3, we show the evolution of
density parameters of various components in universe and the
effective equations of state for DE and DM. The model parameters
are taken to be the same as above. It is exhibited that after a
transient period of baryon domination, the universe enters into a
DE-dominated epoch, and the interaction between DE and DM forces
the ratio between energy densities of DE and DM to remain constant
via DM particles varying mass. In the future, the universe will be
governed by DE and DM which are coupled together, and $\rho_{\rm
b}$ and $\rho_{\rm rad}$ are diluted away with the usual laws,
$a^{-3}$ and $a^{-4}$, respectively. We see that in the attractor
regime the effective equations of state for DE and DM converge
together and both are negative. Notice that the attractor solution
is going to be reached today in this example.

In summary, we investigate in this Letter the attractor solution
of a coupled quintessence scenario in which the scalar field
evolves in a power law potential and the mass of VAMPs depends on
a power law function of $\phi$. It has been shown by numerical
calculation that this interacting scalar field scenario has a
stable attractor and can thus provide a natural solution to the
cosmic coincidence problem.

\begin{acknowledgments}
The author would like to thank Xiao-Jun Bi and Bo Feng for useful
discussions. This work was supported in part by the Natural
Science Foundation of China (Grant No. 10375072).
\end{acknowledgments}


\end{document}